# Model-based pressure tracking using a feedback linearisation technique in thermoplastic injection moulding

Mandana Kariminejad, Centre for Precision Engineering, Material and Manufacturing (PEM), Atlantic Technological University Sligo, I-Form Advanced Manufacturing Research Centre
David Tormey, Centre for Precision Engineering, Material and Manufacturing (PEM), Atlantic Technological University Sligo, I-Form Advanced Manufacturing Research Centre
Marion McAfee, Centre for Precision Engineering, Material and Manufacturing (PEM), Atlantic Technological University Sligo, I-Form Advanced Manufacturing Research Centre

*Abstract*

Injection moulding is a well-established automated process for manufacturing a wide variety of plastic components in large volumes and with high precision. There are, however, process control challenges associated with each stage of injection moulding, which should be monitored and controlled precisely to prevent defects in the injection moulded component. One of the process variables is the pressure profile during the injection and packing phases, which has a direct impact on the quality of the manufactured part. This research proposes a model-based controller design for the injection and cavity pressure during the moulding cycle, with a feedback linearisation controller. First, the injection and packing phases were mathematically modelled and converted to a state-space model. The procedure of designing the controller for the process was outlined. A pressure profile was defined as the target trajectory in the proposed controller and the ability of the designed controller in following the set profile was explored.

## 1. INTRODUCTION

One of the most developed processes to produce plastic components is injection moulding. In general, this process contains three main steps: the filling stage in which melted polymer pellets are injected into the cavity, the packing stage which prevents excessive shrinkage by injecting extra polymer when the cavity is full, and the cooling stage where the polymer solidifies and becomes ready for ejection (Kazmer, 2007). There are process control challenges associated with each stage of injection moulding, critical process variables should be monitored and controlled precisely to prevent defects in the injection moulded component. One of the process variables of interest, is the pressure profile during the packing and injection phases, which has a direct impact on the quality of the manufactured part. Non-optimised pressure will lead to part defects such as weldlines, shrinkage, and warpage (Chen et al., 2019; Kurt et al., 2009).

Real-time and online control of the injection moulding process is a challenge. One of the main challenges is to control and model the batch nature of the process, which is different from classical continuous process models due to the levels of inherent process variation. This process is also nonlinear with a high number of complex and dynamic variables, making the design of a controller more difficult. Several studies have been done to control the process by using a classical controller such as PID (C. J. Chen et al., 2021; Jeong et al., 2015; Pannawan & Sudsawat, 2021). However, these controllers are not able to control the complexity, uncertainties, and nonlinearity of batch processes like injection moulding.

A feedback linearisation controller is a powerful controller which has an acceptable control performance for many nonlinear systems, and it addresses two of the main challenges in control theory: robustness and stability. In this method, a nonlinear system is transformed into a fully or partially decoupled linear system by using nonlinear transformation and cancelling the nonlinearities of the system through feedback. After linearisation, linear control techniques can be applied to control the desired output (Wu & Blaabjerg, 2021). The linear design tool used in this research is input-output linearisation, which can be achieved by differentiating the output several times and has been applied to obtain a linear input-output description.

Email: Mandana Kariminejad
        Mandana.kariminejad@mail.itsligo.ie

In this project, design of a feedback linearisation controller is explored for the nonlinear model of a servo-electric injection moulding machine to control and track the desired pressure profile and find the optimum input, which is the voltage of the servo-electric motor, for the desired pressure profile. In the following sections the nonlinear model of servo-electric injection moulding is outlined and the feedback linearisation method and resulting simulation are discussed. To conclude, a discussion regarding the controller result and further research opportunities for the improvement of the controller are explained.

## 2. METHODOLOGY

### 2.1. Nonlinear model of a servo-electric injection moulding

This study utilises the nonlinear model developed by Stemmler et al. (Stemmler et al., 2019) for a servo-electric injection moulding machine. They modelled the servo-electric drive, plastification unit, nozzle, and cavity. In their model, the servo-electric drive was estimated by a second-order system to relate the input voltage $U$ to the desired drive velocity $v$. The transfer function for this second-order system by the Laplace transformation is presented in equation (1). In the function, the values gain $K=23.4$, damping $D=0.79$ and cut-off frequency $w_0=133s^{-1}$ are estimated.

$$G(s) = \frac{V(s)}{U(s)} = \frac{Kw_0^2}{s^2 + 2Dw_0 s + w_0^2} e^{-sT_d} \tag{1}$$

The plastification unit was approximated as a hydraulic cylinder and the pressure was derived by the mass continuity equation as equation (2). $\beta_s$ is the bulk modulus, $v_s$ is the specific volume and $\dot{m}_n$ is the mass flow through the nozzle and can be also estimated by equation (3).

$$\frac{dp_s}{dt} = \frac{\beta_s}{x}(-v - \dot{m}_n v_s + m_s \dot{v}_s) \tag{2}$$

The mass flow through the nozzle ($\dot{m}_n$) was found by assuming a steady-state flow of a Newtonian fluid. In this equation, the radius and length of the nozzle are $R=0.2$ cm and $L=8$cm respectively. The viscosity $\mu$ is considered constant and equal to 60 kg m$^{-1}$ s$^{-1}$.

$$\dot{m}_n = \frac{\pi R^4}{8 v_s L \mu}(p_s - p_c) \tag{3}$$

The cavity was modelled, similarly to equation (2), by modelling the mass flow through the nozzle and shrinkage of the melt flow in the cavity as described in equation (4). $\beta_c$ is the bulk modulus, and $v_c$ is the specific volume in the cavity.

$$\frac{dp_c}{dt} = \frac{\beta_c}{v_0}(\dot{m}_n v_c + m_c \dot{v}_c) \tag{4}$$

To simplify the model, the derivation of the specific volume of the cavity and screw were neglected. Also, the bulk modulus in both the cavity and screw ($\beta_s, \beta_c$) are considered to be constant and equal to 8662bar. The mathematical model consists of five variables identified as screw position ($x := x_1$), drive velocity ($v := x_2$), derivative of the derive velocity ($\dot{v} := x_3$), screw pressure ($p_s := x_4$) and cavity pressure ($p_c := x_5$). By considering a five-state nonlinear equation the system can be modelled as equation (5), where $x$ is the state vector, $U$ is the input voltage and the desired output $y$ for this system is the cavity pressure which is defined as the fifth state $x_5$ above.

$$\dot{X} = f(x) + g(x)U \tag{5}$$
$$y = p_c = x_4 = h(x) \quad , \quad Q = \frac{\pi R^4}{8 v_s L \mu}$$

where,

$$f(x) = \begin{bmatrix} x_2 \\ x_3 \\ -2Dw_0 x_3 - w_0^2 x_2 \\ -\frac{\beta_s}{x_1} x_2 - \frac{Q\beta_s}{x_1}(x_4 - x_5) \\ \frac{Q\beta_c}{v_0}(x_4 - x_5) \end{bmatrix}, \quad g = \begin{bmatrix} 0 \\ 0 \\ kw_0^2 \\ 0 \\ 0 \end{bmatrix}, \quad x = \begin{bmatrix} x_1 \\ x_2 \\ x_3 \\ x_4 \\ x_5 \end{bmatrix}$$

## 2.2. Feedback linearisation methodology

The method used here is based on single input feedback linearisation developed by (HASSAN K. KHALIL, 2002). By considering the following SISO (single input, single output) system and having the Lie derivative of $h$ along the direction of the vector $f$:

$$\dot{X} = f(x) + g(x)U \tag{6}$$
$$y = p_c = x_5 = h(x), \quad L_f h := \frac{\partial h(x)}{\partial x} f$$

The derivative of output ($\dot{y}$), can be written by equation (7).

$$\dot{y} = \frac{\partial h}{\partial x}(f(x) + g(x)U) \stackrel{\text{def}}{=} L_f h(x) + L_g h(x)U \tag{7}$$
$$L_g h := \frac{\partial h(x)}{\partial x} g$$

In the injection moulding model based on equation (5), $L_g h(x) = 0$. So, the derivative of the output should be further differentiated until $U$ appears in the equation. In our model after four derivatives, this condition was satisfied. The relative degree of the system is 4, where the condition of $L_g L_f^{4-1} h(x) \neq 0$ was satisfied.

$$y^{(4)} = L_f^4 h(x) + L_g L_f^{4-1} h(x) U \tag{8}$$
$$U = \frac{1}{L_g L_f^{4-1} h(x)}[-L_f^4 h(x) + v], \quad y^{(4)} = v$$
$$L_f^4 h(x) = L_f L_f^{4-1} h(x), \quad L_g L_f h(x) = \frac{\partial L_f h}{\partial x} g$$

In our model:

$$L_g L_f^{4-1} h(x) = \frac{KB_s w_0^2}{x_1} \tag{9}$$
$$L_f^4 h(x) = \frac{Q^2 B_s^2}{v_0}\left(\frac{x_3}{x_1^2} - \frac{2x_2^2}{x_1^3}\right) + \left(\frac{Q^3 B_s^3 B_c}{v_0 x_1^3} + \frac{Q^2 B_s^2 B_c}{v_0^2 x_1^3} x_2 + \frac{Q^3 B_c^2 B_s^2}{v_0^2 x_1^2} x_2 + \frac{Q^3 B_c^3 B_s}{v_0^3 x_1}\right)(-x_2 - Q(x_4 - x_5)) +$$
$$\left(\frac{2Q^3 B_s^2 B_c}{v_0 x_1^3} + \frac{Q^3 B_c^2 B_s}{v_0^2 x_1^2}\right)(x_2 x_5 - x_4 x_2) - \left(\frac{Q^4 B_s^3 B_c^2}{v_0^2 x_1^2} + \frac{Q^4 B_c^3 B_s}{v_0^3 x_1} - \frac{Q^4 B_c^4}{v_0^4}\right)(x_4 - x_5) + \frac{Q^2 B_s B_c}{v_0^2 x_1^2}\left(x_4 x_3 - \frac{x_2^2 x_4}{x_1}\right) +$$
$$\frac{Q^2 B_c^2 B_s}{v_0^2 x_1}\left(x_3 - \frac{x_2^2}{x_1}\right) - \frac{B_s}{x_1}\left(-2Dw_0 x_3 - w_0^2 x_2 - \frac{x_2 x_3}{x_1} + \frac{2x_2 x_3}{x_1} - \frac{x_2^3}{x_1^2}\right)$$

By considering a feedback controller it can be assumed that $y^{(4)} = v = -Ke$, where system error $(e)$ is equal to $y - y_d$ and $y_d$ is the desired output and thereby pressure profile. By considering a feedback controller it can be written:

$$v = y^{(4)} = y_d^{(4)} - k_1 \dddot{e} - k_2 \ddot{e} - k_3 \dot{e} - k_4 e \xrightarrow[y^{(4)} - y_d^{(4)} = 0]{} k_1 \dddot{e} + k_2 \ddot{e} + k_3 \dot{e} + k_4 e = 0 \tag{10}$$

where $k_1, k_2, k_3$ and $k_4$ are the controller gains and should be estimated and adjusted so that the controller successfully tracks our desired pressure profile.

## 3. RESULTS

MATLAB R2021 Simulink was used to model the controller and the system. The Simulink model consists of three functions. The first block defines the model of the injection moulding machine where the cavity pressure is the output. The second is the feedback linearisation control strategy as explained in the methodology section and the last one creates the feedback gains and error. The schematic of the simulation model is presented in Figure 1.

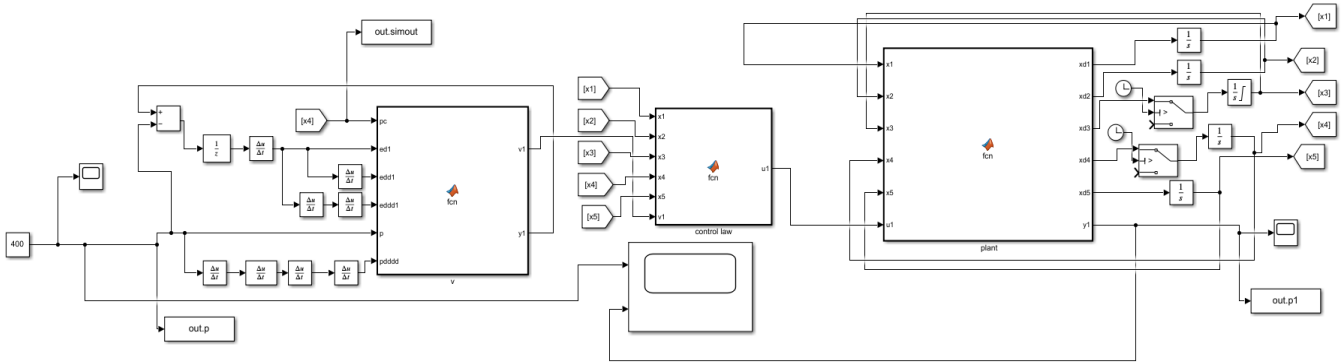

*Figure 1. Simulink simulation of the system and controller*

The performance of the controller when the cavity pressure is constant and equal to 400bar was investigated and presented in Figure 2. After almost five seconds the cavity pressure reached the desired set point of 400 bar.

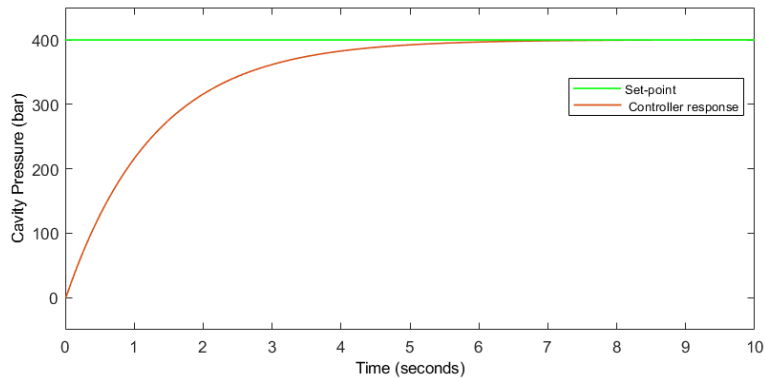

*Figure 2. Feedback linearisation response with constant cavity pressure set point*

The controller gains were chosen as $k_1 = 0.7$, $k_2 = 2$, $k_3 = 30$ and $k_4 = 2.5$.

## 4. CONCLUSION

In this paper the nonlinear mathematical model of a servo-electric injection moulding process is presented. The model contains five states named screw position, derive velocity, derivative of derive velocity, screw pressure and cavity pressure. The model was used in a feedback linearisation control approach to track a constant cavity pressure profile. The controller was designed and simulated in MATLAB R2021 Simulink.

The controller successfully tracked the cavity pressure profile while the response time was slow for the injection moulding process. The response time of the controller depends on the controller gains ( $k_1, k_2, k_3, k_4$) which should be optimised to provide the best performance. These gains can be found and optimised through machine learning algorithms.

In this research, the performance of the controller was only investigated against the constant cavity pressure profile, further research is required to evaluate the controller response with other pressure profiles and other desired outputs such as screw position and pressure. In this study the gains of the controller were selected randomly, for future work these gains should be optimised through a machine learning algorithm. Finally, the controller was designed based on the continuous model, however, injection moulding is a batch process and for better estimation, the controller should be designed based on the discrete model to have more reliable and realistic results.